# DNA Methylation in hypoxia in *Mycobacterium tuberculosis*

**Nayada Pandee[1], Prasert Auewarakul[2], Chanati Jantrachotechatchawan[2,*],**

[1] Triam Udom Suksa School, Thailand;
nayada2248@gmail.com
[2] Research Department, Faculty of Medicine Siriraj Hospital, Mahidol University
chanati.jan@mahidol.ac.th

[*] Corresponding author

## Abstract

Tuberculosis is one of the most lethal contagious diseases caused by *Mycobacterium tuberculosis* (MTB), in many cases, the infected did not show any symptoms, because the bacilli entered the dormant stage in granulomas. The dormant stage of MTB is also associated with higher resistance to drugs and the immune system. Among multiple epigenetic regulations critical to MTB stress responses, DNA methylation is necessary for the survival of MTB in hypoxic conditions, which is a common stress event during granuloma formation. This review gathers previous findings and demonstrates a meta-analysis by collecting hypoxia gene expression data from several articles and perform association analysis between those genes and methylation site profiles across whole genomes of representative strains pf lineage 2 and 4. While more data is required for more conclusive support, our results suggest that methylation sites in the possible promoter regions may induce differential gene regulation in hypoxia.

## Introduction

Tuberculosis (TB) is an airborne, fatal disease caused by infection with *Mycobacterium tuberculosis* (MTB), and it was a epidemic disease that was widespread in 2019. On that occasion, World Health Organization announced that Thailand was in the TOP 14 countries with a tuberculosis burden (WHO). Some people have been infected with tuberculosis for decades before it is prognosticated, and some people have been infected with tuberculosis and never show symptoms during life. After inhaling air with MTB, MTB approach the trachea and then goes to the lung surface, immune cells in the host attack and phagocytose pathogenic bacteria. Immune cells congregate in response to MTB and form granulomas which means that the immune cells can not eliminate the MTB but can inhibit its dissemination. This state where MTB survive but are completely contained by immune structures e.g. granulomas is called the latent tuberculosis infection (LTBI). In the center of granuloma, oxygen depletion hinders MTB growth and replication. During hypoxia condition, MTB has a lot of epigenetic modification to survive and increase the probability of breaking out from granulomas. Epigenetic adaptation of MTB plays a significant role in evading host immune and undermining cellular defense mechanisms of macrophages. Understanding epigenetic modification during hypoxia is important for TB diagnosis and remediation (reviewed by Sui et al. 2022) . This review discusses the potential relevance of DNA methylation within the MTB cells under hypoxia conditions and conducts meta-analysis to support the hypothesis that DNA methylation may influence gene expression during hypoxia.

## DNA methylation in *Mycobacterium*

DNA methylation is one of the epigenetic controls that modify gene expression to be suitable for the environment. DNA methylation refers to the process of adding a methyl group to an ordinary standardized base in DNA. It does not cause any changes in genetic sequences but can add information that helps regulate the expression of genes. S-adenosylmethionine (SAM) can be the methyl-group donor, then DNA methyltransferase (DNMT) carries a methyl group through covalent bonding and eventually transfers to the nitrogenous base (Song et al. 2020). Methylation in promoters decreases gene expression mainly in eukaryotes, while methylation in gene sequence may directly increase gene expression. Methylation occurs in DNA major groove and does not interfere with hydrogen bonds between base pairs. One of the most widely studied methylations after 5-methylcytosine is $N^6$-methyladenine, known as 6mA in DNA and $m^6A$ in RNA. The Amounts of $m^6A$ and 6mA are dynamic, but abnormalities in methylation rates are associated with some diseases.

### Detection of 6mA and $m^6A$ by nitrite-mediated deamination

The effort to detect methylation has a long history, but all methods have some limits. Immunoprecipitation was one of the popular methods. However, traditional IP-based methods, such as 6mA-DIP-seq provide low resolution. Although IP methods coupled with restriction digest, such as DA-6mA-seq have improved resolution, they are only efficient for short sequences. PacBio single molecule real-time (SMRT) sequencing technology provides high resolution through the single-nucleotide level, but it has a lot of false positives, does not deal with long sequences, and is not suitable for 5mC. Another method is 6mA-crosslinking-exonuclease- sequencing (6mACE-seq) provides high resolution to a single-nucleotide level, but the process is cumbersome.

The nitrite-mediated diazotization of aromatic amines, first described by Griess, is the method that can be applied to detect methylations. This method has four advantages, this reaction is water resistant, chemoselective for either N6 -methyladenine or unmethylated adenine, the reaction does not degrade DNA or RNA sequence, and this reaction brings about a change in how polymerase or reverse transcriptase represents the result when reading nucleobase. In acidic conditions, nitrite turns to reactive nitrosonium ion which reacts with aromatic amines and produces nitrosamine, then dehydrated to form the diazonium ion. These processes mentioned are only found in primary aromatic amines, not in secondary aromatic amines due to the lack of available electron pairs for the elimination of water and triple bond formation. Only unmethylated adenine will be hydrolyzed under these conditions to form hypoxanthine (a nucleoside with hypoxanthine base is called inosine) so that primary exocyclic amine of adenine and exocyclic amine of N6-methyladenine can be differentiated. Treatment with $NaNO_2$ in the presence of acetic acid (AcOH) over a 12 h period results in a complete conversion of adenosine into inosine, and deamination of guanosine into xanthosine and cytidine into uridine. Deamination of adenosine into inosine was over 50% completed within 5 h without damage to DNA or RNA, so a 5 h period is sufficient to probabilistically detect the difference of deamination at methylated sites by next-generation sequencing. Surprisingly, 6mA/$m_6A$ becomes nitrosylated significantly faster than adenosine possibly owing to its risen nucleophilicity as $N^6$ becomes a secondary amine. The concentration of AcOH also determines reaction efficiency as 5% AcOH for RNA and 2.3% AcOH for DNA represented the best conditions for deamination activity (Mahdavi-Amiri et al. 2020).

## Methylation in *Mycobacterium tuberculosis* during infection to host cells

There are three DNA methyltransferases (DNA MTases) that are associated with three different 6mA methylation motifs found in homologous genes of every strain of MTB (Zhu et al. 2016). These DNA MTases are MamA, MamB, and HsdM. DNA MTase in prokaryotes works under the R-M system. MamA is efficient in the Euro-American strains (L4) i.e. H37Rv, while it is inactivated in the Beijing strains (L2) due to the loss of function mutation E270A. *lppC* gene is said to be protected from cleavage by PvuII in MTB strains from the Euro-American lineage (L4), while Beijing strains showed lessened *lppC* protection than others (Shell et al. 2013). Furthermore, deletion of *Rv3263*, which encodes MamA, in H37Rv (L4) also results in a lack of *lppC* preservation. This result supports that MamA is the methyltransferase that protects *lppC* gene from endonucleolytic cleavage. PvuII recognition sites are present in every strain of MTB, while methylation affects what type of cleavage will be. PvuII recognition site is not disrupted by the cleavage, and methylation of adenine within the recognition site is predicted to obstruct PvuII cleavage. MamA seems more like an orphan MTase, which functions in chromosome stability, mismatch repair, and replication. MamB and HsdM are also known as DNA MTases and both of them are predicted to encode IIG type MTase and type I MTase. Interestingly, HsdM seems more conserved than orphan MTase. Type II DNA methyltransferases (MTases) are ubiquitous in prokaryotes. They have main functions such as host protection and epigenetic regulation. The types of MTases are categorized into class I which forms N-6-methyladenine, class II which forms N-4-methylcytosine, and class III which forms C-5-methylcytosine. MTases that are mentioned are relevant to restriction endonuclease (REase) and are called the R-M system which protects bacterial cells from invasion by foreign DNA. Five methylation-affecting mutations are found, and probably disrupt gene methylation. Some strains in lineage 1 have S253L mutation in MamB. The *mamA* E270A is found in every strain in lineage 2. Meanwhile, all strains in lineage 3 have G173D and L119R methylations at HsdM and HsdS, respectively. Additionally, more than half of the strains in lineage 4 have P306L mutation in HsdM.

| MamA | MamB | HsdM |
|---|---|---|
| **5′-CTCCAG-3′**<br>3′-GAGGTC-5′ | **5′-CACGCAG-3′**<br>3′-GTGCGTC-5′ | **5′-GATNNNNRTAC-3′**<br>3′-CTANNNNYATG-5′ |
| **5′-CTGGAG-3′**<br>3′-GACCTC-5′ | **5′-CTGCGTG-3′**<br>3′-GACGCAC-5′ | **5′-GTAYNNNNATC-3′**<br>3′-CATRNNNNTAG-5′ |

**Figure 1.** Methylation motif for each of the 3 major DNA MTases in MTB. Bold indicates main genomic reading strand and red highlights an adenine base methylated by a given enzyme.

**Table 1.** This table illustrates the specific DNA MTase mutations of each major MTB lineage and its representative strains. L5 and L6 are not included since most of their strains do not have mutations in MamA, MamB, or HsdM.

| | Lineage 1 | Lineage 2 | Lineage 3 | Lineage 4 |
|---|---|---|---|---|
| mutation | *mamB* S253L | *mamA* E270A | *hsdM* G173D<br>*hsdS* L119R | *hsdM* P306L |
| examples | East African - Indian (EAI) | Beijing | Dehli-CAS | Cameroon<br>H37Rv<br>Haarlem |

# Hypoxia

*Mycobacterium tuberculosis* (MTB) is an aerobic bacteria that needs sufficient oxygen to grow and complete its life cycle. Hypoxia is an oxygen-depleted condition. During MTB infection, macrophages congregate, surround MTB, and form granulomas. In the center of granuloma, there is oxygen depletion, called hypoxia. When MTB is completely covered by leukocytes, MTB can not grow and replicate efficiently because lack of sufficient oxygen levels. On the other hand, MTB can cause multiple dysfunctions including microRNA systems in the infected and surrounding macrophages alike i.e. miR146a is overexpressed, resulting in suppression of TRAF6, in the host cells to increase MTB's chance of survival. MTB is dormant in this stage, but ready to erupt from granuloma. Identification of pathological stages or disease progression from latent to active TB infection in patients is a concerning problem. Multiple models have been developed to understand the actual conditions and infection mechanisms of MTB in the human host tissue such as epigenetic changes during hypoxia and the resurrection of MTB from the dormant stage induced by hypoxia.

Wayne's dormancy model illustrates hypoxia-induced dormancy and proves that oxygen depletion can induce MTB to be dormant *in vitro* similar to hypoxia and acidic in granulomas (Wayne et al.). Latent MTB complex strains are detected by changes in growth, metabolic activity, and transcriptional profile (Tizzano et al. 2021). The shift down to dormancy can be divided into three stages. In non-replicating persistence stage 1 (NRP-1), the rate of increase in turbidity is slowed down, and no increase in turbidity in NRP-2. The next stage is a viable but non-cultivable state (VBNC), which is found in the H37Rv strain after has been dormant for three to four weeks, main replicative and metabolic activities are shut down. Tizzano et al. (2021) assumed that dormant MTB and dead MTB can be separated by providing oxygen and following up if they regain metabolic activity and the ability to grow and divide. Only MTB of lineage 4 (L4), but not those of L1 and L2, can resurrect upon oxygen exposure after a long hypoxia-induced dormancy. The reactivation rate measured as OD594 of the liquid medium of Beijing after 7 days of oxygen starvation surpasses that of H37Rv on the 11th day of reactivation, but the CFU result shows that the Beijing reactivation rate is still less than H37Rv. Nevertheless, both types of the experiment show an efficient growth of Beijing and H37Rv. When the hypoxia period was extended to 14 days, the H37Rv reactivation rate still increased normally, while Beijing can rarely resurrect. Furthermore, when the researchers prolonged the hypoxia period to 21 days and 35 days respectively, only H37Rv can revive. Besides H37Rv and Beijing, which are the representative strains for lineage 2 and 4 respectively, East African - Indian (EAI) representing lineage 1 and Delhi/CAS representing lineage 3 were also included in the experiment. Additionally, researchers add Cameroon and Haarlem from lineage 4 to this study along the common lab strain H37Rv. The result shows that the overall reactivation trends of Haarlem, and Cameroon resurrection are similar to H37Rv, while reactivation of EAI, Delhi/CAS are similar to Beijing. It can be seen that the reactivation mechanisms are lineage-specific. Specifically, lineage 4 has a special characteristic to endure prolonged hypoxia up to 35 days in the experiments. Loss of metabolic activity during oxygen starvation was assessed by esterase-dependent hydrolysis of fluorescein diacetate (FDA) via flow cytometry. However, this research use only oxygen to reactivate the dormant states without applying any other resuscitation factors. Several studies (Dusthackeer et al. 2019; Chengalroyen et al. 2016) have shown that a long period of can dormancy can makes MTB lose efficacy to form colonies in solid media, but it is still capable to resurrect in liquid media after spontaneous reintroduction of oxygen or gain proteinaceous reactivation factors. As discovered by Chengalroyen et al. (2016), Beijing requires RPF to recover from hypoxia and still requires unknown chemical agents from the culture filtrate of active MTB. Beijing also produces rpfA more than H37Rv approximately 2.6 folds during hypoxia, Tizzano et al. (2021). The above findings together with a hypoxia-associated increased production of glbN hemoglobin protein in Beijing compared with H37Rv support the notion that Beijing may enter a deeper and mechanistically different hypoxia-induced dormant condition compared with H37Rv and other L4 strains.

During hypoxia, MTB-infected macrophages assemble lipid droplets which are advantageous for dormant MTB in host cells. Therefore, Oil Red-O stained tags lipid droplets and illustrates a significant increase in signal. Acid-fast staining can detect MTB during normoxia because it tags mycolic in the MTB cell wall, while there is a mycolic derivative during hypoxia that causes MTB to no longer be detected by acid-fast staining but can be detected by Nile Red instead (Daniel et al. 2011).

## Methylation During Hypoxia

Under the hypoxia condition, MTB has to modify its genetic material to be agreeable to survive. According to Tizzano et al. (2021), researchers found the difference between Beijing (L2) and H37Rv (L4) expression during oxygen depletion. Researchers harness the NRP-2 stage after 10 days of O2 starvation to compare transcriptome analyses of Beijing (L2) and H37Rv (L4). A strong upregulation of stress- and oxygen starvation-related genes were detected in H37Rv. There were 471 upregulated and 483 downregulated genes found in H37Rv.

### Mycobacterial adenine methyltransferase (MamA)

MamA is a significant DNA methyltransferase that is encoded by *Rv3263* and functions in providing $N^6$ methyladenine. Shell et al. (2013) discovered that MamA influences gene expression and contributes to a crucial, but strain-specific, regulation especially during hypoxia. Δ*mamA* strain died significantly faster than the wildtype in an absence of oxygen. According to Shell et al. (2013), there is a group of genes that were considered to be directly affected form DNA methylation, including *Rv0102*, *Rv0142*, *corA*, *whiB7*, and the *Rv3083* operon, this hypothesis was rechecked by using quantitative PCR, the result confirmed that lacking MamA leads to a significant decrease in *Rv0102*, *Rv0142*, *corA*, *whiB7* expression. *Rv0102* predictedly encodes an uninvestigated integral membrane protein and can be moderately induced by oxidative stress; *Rv0142* is expected to encode DNA glycosylase and is strongly activated by oxidative stress, additionally, *CorA* encodes a predicted magnesium and cobalt transporter, and can be moderately induced by thioridazine, proton gradient disrupters, and oxidative stress. Although MamA methylation is important to *Rv0142* and corA transcription at the basal level, the down-regulated transcription from the lack of MamA can be compensated by transcription up-regulation in response to oxidative stress. *WhiB7* is a transcriptional regulator that can be induced by several stress factors. To find the transcriptional start sites (TSS) of these genes to understand the real mechanism of MamA, 5′ and 3′ ends of each mRNA were joined together by using T4 RNA ligase RNA in order to circularize the samples and enable amplification and sequencing of 3′ tail and 5′ head of the mRNA in a single segment. The circular templates are applied to synthesize linear cDNA by random priming. The reverse primer was set to pair with the sequence that locates a little bit after the start codon, while the forward primer was designed to match with the sequence near 3′ ends, this strategy resulted in the production of amplicon containing 5′-3′ junctions. The TSSs of *Rv0102*, *Rv0142*, *corA*, *whiB7* were discovered that they each locate on four to five base pairs downstream of a MamA methylation site. Moreover, it was shown that the putative sigma factor -10 site overlaps with MamA recognition site. Specifically, the last base, here in A, on the template strand of the sigma factor -10 site will be methylated and the third base on the coding strand downstream to the sigma factor -10 site, also A, will be methylated as well. However, the overlap between a methylation site and the sigma factor -10 site was first discovered in E.Coli in the case of *dnaA*, IS10 transposase, and *glnS*. Moreover, hemimethylated promoters and complete methylated promoters differentially affect gene expressions, mostly involved in the cell cycle. Deletion of *ΔmamA* does not significantly change growth rate compared with that of wildtype H37Rv under standard growth conditions and mouse infection. Additionally, *ΔmamA* deletion does not affect overall mutation rates. However, there is one condition that is more obviously found in humans, hypoxic granulomas. *ΔmamA* death rate was significantly high, while other strains present a drop in viability. Tizzano et al. (2021) have identified genes differentially regulated between H37Rv standard condition vs 10-day hypoxia and between H37Rv vs Beijing during their 10-day hypoxia. While methylation by MamA is essential to the survival of H37Rv during hypoxia, none of the 3 major MTases namely MamA, MamB, and HsdM was found significantly changed in the experiments by Tizzano et al. (2021). Nonetheless, this is congruent with a methylation profile study by Zhu et al. (2016) that 99.7% of MamA methylation site in H37Rv are methylated under a standard growth condition, leaving only 1 intergenic motif, which may be involved in transcriptional regulation, unmethylated. MamB methylation rate is either 0% in H37Rv or 100% in Beijing. Interestingly, HsdM methylation rate in Beijing is approximately 97% and the remaining sites are hence possibly methylated under stress conditions.

## DosR Regulon

DosR regulon is a specific genetic program used in MTB and is proven to be involved with the survival of MTB while incapable to continue the aerobic activity, including hypoxia, excessive amounts of NO, or CO conditions that trigger DosR regulon's function. DosR has three main constituents, DosS and DosT are histidine kinases that function as the sensor of NO and CO, while DosR takes part in response or is called a response regulator. DosR regulon regulates at least 48 genes downstream. Several research findings discovered that DosR regulon involves MTB survival during latent infection. Leistikow et al. (2010) demonstrated the growth rate of wildtype MTB compared with DosR regulon mutant under hypoxia. H37Rv represents the wild type, DorKO is H37Rv with *Rv3134c*, *Rv3133c*, and *Rv3132c* deleted, and DorCO is DorKO complemented with *Rv3134c*, *Rv3133c*, and *Rv3132c* via an integrative plasmid, with the native promoter sequence intact. By studying based on Wayne's model, the Rapid anaerobic dormancy model (RAD) was introduced. This model shortened the time by using larger stir bars, rapid stirring, and sealed the test tubes with grease. Under the condition of oxygen depletion, the growth and survival rates of the wild type, DorKO, and DorCO were monitored. Overall it is clear that all three types of samples decreased, DorKO entered the nonreplicating stage within a few days and had significantly dropped faster than any type, while H37Rv and DorCO shared a similar pattern of a reduction of growth rate (CFU/ml) which entered the stage in which no additional growth by day 7. The result from the MPN technique in liquid medium and comparison to CFU values on plate also supported the first result. DosR regulon plays an important role in the survival of MTB during a lack of oxygen. Losing DosR regulon undermines the survival of MTB. Moreover, dormant MTB is nonculturable in solid medium except in liquid medium. Leistikow et al. (2010) revealed that up to day 20, the mutants presented metabolic activity but most of them could not recover on solid media; however, they all died at last. Additionally, DosR regulon mutant MTBs tend to take oxygen more rapidly than the wild type. DosR regulon also takes part in ATP levels maintenance, in the beginning, ATP levels rise as the number of cells increases, after that by day 5 DosKO' ATP levels dropped dramatically compared with day 4 and became declined moderately and remain lower than the wild type. Similar to ATP levels modification, NAD levels dropped after 5 days of oxygen starvation. However, the NAD/NADH ratio gradually declined for DosR regulon mutant, but this ratio in the wild type had fallen then went up again to around 6 times compared with the mutant, so these findings supported that DosR regulon also plays a part in redox balance. Reintroduction of oxygen after 10 days of dormancy showed the general overview that DorCO recovered fastest which grew to approximately 16-fold increase of CFU from Day 0 by 5 days after having abrogated oxygen starvation. This is followed by H37Rv and DorKO, with the figures being at about 12 and 8-fold increase of CFU from Day 0 by 5 days, respectively. Resurrection after 20 days of dormancy, by 5 days H37Rv reached approximately 12 fold, come after by DorCO at nearly 10 fold increase of CFU from Day 0, this is opposed to DorKO which the resurrection rate remains stable at 0 fold increase from day 0 to day 5. DosR regulon expression profile can differentiate Beijing strains from non-Beijing strains. Domenech et al. (2016) demonstrated this assumption by using HN878 to represent the modern Beijing strain, 98_1663 to represent the ancestral lineage 2 strain which lacks the *RD207* deletion, and H37Rv to represent the non-Beijing strain. After incubating these three strains under standard *in vitro* conditions, the result showed that the ancestral Beijing strains share a similar pattern of gene expression. Beijing strains lack the function of MamA, while DosR regulon plays a crucial role in surviving hypoxia. Moreover, DosT, one of the sensors in the DosR regulon lacks function in this lineage; therefore DosR is constitutively stimulated as confirmed by comparative real-time PCR analysis. Three synonymous SNPs located nearby *dosR* including *Rv3131*, *dosS*, and *Rv3134c* were reported. Two of them are *C507G* and *C601T*, the former presents in both the ancestral East Asian and modern Beijing strains, while the latter was only found in the modern Beijing strains. The *C507G* SNP is specific to East African/Indian, while the *C601T* SNP is a concrete characteristic of the Beijing sublineage of the East Asian lineage. Domenech et al. (2016) further illustrated the significant evidence by using phage engineering to contain a wide-type non-Beijing version ground on the H37Rv sequence (pNS series) and a Beijing version by using the sequence of HN878. The expression of *dosR* was monitored by using quantitative real-time PCR (qRT-PCR) and utilized the *sigA* gene to encode the housekeeping sigma factor. This experiment resulted in the outstanding effect of the Beijing-specific sequence on dosR expression even if placed in a non-Beijing strain (H37Rv) (Domenech et al. 2016).

## Meta-analysis Methods

Differential gene expression data between normal condition and 10-day hypoxia H37Rv and between H37Rv and Beijing type 1500/03 strain both at 10-day hypoxia are collected from Tizzano et al. (2021). Differential gene expression data between H37Rv and Beijing type HN878 strain both at baseline normal condition are reported by Domenech and colleagues (2017). From the 3 sets of data above, differential expressions between normal condition and hypoxia in Beijing type strains were inferred under an assumption that genes not reported showed no difference or lower than 2 folds change in expression. Lastly, the differential regulation upon hypoxia between Beijing vs H37Rv (r.B/R) was calculated with lower and upper bounds for low fold changes set to 0.667 ($\log_2(1/1.5) = -0.585$) and 1.5 fold ($\log_2 1.5 = 0.585$) respectively. Genes showing r.B.R with lower bound changes of at least 2 folds were selected for further discussion. Since genes in DosR regulon are upregulated during hypoxia, the list of DosR regulon genes

Whole genome sequences and gene coding sequences (CDS) of L2 representative Beijing type strain Mtb 2279 (GenBank: CP010336.1) and L4 representative clinical isolate CDC1551 (GenBank: AE000516.2) were downloaded from NCBI. To match the genomic sequence with the differential expression results from Tizzano et al. (2021) and Domenech et al. (2016), microarray probe sequences based on genes of H37Rv (eArray design ID 064078) were mapped to the CDS of the two representative strains. The cutoff of at least 90% overlapping alignment score was chosen considering that all 8 genes of Mtb 2279 that passes 90% but not 95% still have matching gene description with the H37Rv probe target genes. Out of 4,599 and 4,189 genes genes from Mtb 2279 and CDC1551 respectively, a total of 3,556 genes overlap between the two representatives.

Annotation of operons across the genes of CDC1551 was conducted by Pelly and colleagues (2016) using experimental data from RNA sequencing. The corresponding 3,556 overlapped genes between the 2 strains were assigned operons based on the CDC1551 data. The orientation of each gene was confirmed to be the same between the 2 strains. The distances between consecutive genes within operons were measured and compared between the two strains using paired Student's *t*-test.

The motifs of the 3 main MTases are nearly but not palindromic. Hence, the non-overlapped forward and reverse sequences of each motif are mapped to the genomes of the representative 2 strains independently. The locations are designated as either in the coding region (mt_c) or in the upstream promoter-associated region (mt_p). The methylation sites with starting nucleotide downstream to the first base of start codon and upstream to the last nucleotide of coding sequence are denoted mt_c. To be denoted promoter-associated site or mt_p, the sequence must be in the non-coding regions and at most 200 nucleotides upstream to the start codon of the corresponding gene or to a start codon of any upstream genes in the same operon. This rule was applied to include any potential alternative transcriptional start sites (TSS) within an operon. For example, *whiB7* or *Rv3197A* or *MT3290.1* is the last gene in the operon of 5 reverse genes [5′– *MT3294 – MT3293 – MT3291 – MT3290.2 – MT3290.1* –3′]. REMap predicted a full operon during an exponential growth phase, but a sub-operon of MT3291-MT3290.1 during a stationary phase. Interestingly, while there is no MamA methylation site in the upstream promoter regions of both REMap-predicted operons, there is one MamA site upstream to the MT3290.2 gene and this site has been experimentally confirmed to overlap with an alternative promoter of *whiB7/MT3290.1*-containing sub-operon that is expressed during hypoxia (Shell et al. 2013).

The distribution of coding and promoter-associated methylated sites of the 3 MTases across genes grouped by differential expression between the relevant pairs of strain conditions was analyzed using chi-square test via *chisq.test* function in base R. To illustrate, L4 MamA methylation site distribution profile will not be calculated with respect to the genes differentially regulated between normal condition and hypoxia of the L2 strain. For each strain, the resulting chi-square *p*-values were adjusted using false-discovery rate (FDR). The second set of chi-squared analysis was to test the hypothesis whether genes in the same operon, specifically those without upstream methylation sites and those with upstream sites show similar or differential expression.

To compare the differential regulation of genes with or without promoter-associated MT sites within each same operon, sums of log2 fold changes (log2fc) were calculated for the two groups within each operon and then compared across all operons using paired Student's *t*-test. An alternative evaluation is to assess whether presence or lack of MT sites in the promoter region are effective as a clustering criteria. For this task, Calinski-Harabasz (CH) index or variance ratio criterion is used as an internal cluster validation. Herein, CH index of genes clustered by having or lacking promoter-associated MT sites in each operon is compared with the median of CH indices of all combinations of two groups using paired *t*-test.

## Meta-analysis Results and Discussion

### *Differential expression data analysis*

According to data analysis using information from Tizzano et al. and Domenech et al., There are 47 genes have a significant increase in Beijing in hypoxia from the baseline (genes expression in standard condition) compared with the increase of gene expression in H37Rv under the hypoxic condition from gene expression of H37Rv within the standard condition, we termed this value as r.B/R (Table S1). The striking point is that *Rv3862c* (*whiB6*) shows the highest r.B/R. value at approximately 6.509834 folds. The WhiB is a family of redox sensor proteins. *whiB6* expression is upregulated during NO treatment, infection of macrophages, and prolonged hypoxia. Additionally, WhiB6 is involved in mycobacterial infection, granuloma formation, contagion, and disease development. *whiB6* encoded WhiB6 which regulates the expression of genes associated with the ESX-1 secretion system (Table S2). (Chen et al. 2016) The ESX system is used by *Mycobacterium* to transpose protein substrates through the cytoplasmic membrane. Moreover, the ESX system plays an important role in the virulence of *Mycobacterium tuberculosis* (Bosserman et al. 2017). According to Abdallah et al. (2019), using transcriptome sequencing, the result showed that lacking ESX-1 function in both *Mycobacterium marinum* and *Mycobacterium tuberculosis* results in the escalation of the espA operon expression during the infection of macrophages, and another important result they found is that the disturbance of ESX-1-mediated protein secretion causes the decrease of the ESX-1 substrates excepted structural components of the ESX-1 secretion system in culture medium, so it was reported that down-regulation of ESX-1 substrates is caused by a regulatory process influenced by *whiB6* (Abdallah et al. 2019). The WhiB6 Fe-S cluster is essential for the ESX-1 function and the transcriptional profile in Mm also relates to this Fe-S cluster. WhiB6 reacts with NO to dynamically regulate the ESX-1 and Dos dormancy genes and isoforms of WhiB6 also tune the virulence of Mm and take part in granuloma formation in zebrafish. (Chen et al. 2016)

The *Rv0083* is one of the DosR regulon genes which is probably an oxidoreductase, and it becomes the second lead in our data analysis result. However, the information on this specific gene is still limited. The following gene, *Rv3829c* which is probably the dehydrogenase (not in DosR regulon) also lacks data.

Among the last three genes, which have the lowest r.B/R. value, the *Rv2816c* is the one gene for which background information of this gene is already provided. Rv2816c is the *Mycobacterium tuberculosis* CRISPR-associated Cas2 which functions as an endoribonuclease. Among pathogenic mycobacteria such as *M. tuberculosis* H37Rv, *M.*

*tuberculosis* H37Ra, *M. africanum*, *M. bovis*, *M. canetti*, Cas2 is highly conserved. Besides, *Rv2816c* transferred from *M. tuberculosis* is also able to express completely in *M. smegmatis* even though *M. smegmatis* does not have a CRISPR-Cas system. Additionally, *Rv2816c* probably regulates some of the associated genes that cause the variation of components of *M. smegmatis* cell envelope thereby causing the changing of the morphology of the recombinant bacilli. However, Huang et al. (2016) demonstrated that *Rv2816c* recombinant *M. smegmatis* is more vulnerable to ofloxacin and norfloxacin, and *Rv2816c* results in a decrease in the survival rate of *M. smegmatis* within THP-1 cells due to bacterial physiology (cell wall) that was changed. Overexpression of *Rv2816c* also takes part in different stress responses, Rv2816c down-regulates the transcription of *M. smegmatis* stress-related genes (Huang et al. 2016). Rv1576c is the probable PhiRv1 phage protein, and the last one, Rv0073 is the probable glutamine-transport ATP-binding protein ABC transporter.

### *Genes and methylation sites association*

Although there is no confirming RNA-sequencing data or operon annotation on Lineage 2 strains, all of the 3,556 matched genes have the same orientations across both strains. Nonetheless, paired *t*-test across 338 operons between the 2 strains demonstrated that Mtb 2279 has significantly higher mean (mean difference = 84.45; *p*-value = $9.409 \times 10^{-13}$) and higher standard deviation (mean difference = 105.46; *p*-value = $1.135 \times 10^{-13}$) of distances between consecutive genes within an operon, supporting a notion of potentially more complex regulatory systems in Beijing type strains.

When all genes are labeled as either having methylation site or not, distribution of any methylation site types do not significantly differ across groups of genes differentially expressed in any pairs of conditions after *p*-value adjustment. The only significant results by raw *p*-values are CDC1551 coding-region MamA sites (cd mt1.c) and Mtb 2279 coding-region MamA sites (be mt1.c) with respect to the differential expression between normal condition and hypoxia of their respective lineage (R1/R0) ($p = 0.0050 \rightarrow 0.1504$) and (B1/B0) ($p = 0.0067 \rightarrow 0.2012$). The major observed associations are lower number of site-containing and higher number of site-lacking genes in the group of downregulated genes (see Pearson residuals in supplementary Table S3). The remaining significant association is higher MamB site-containing genes with genes relatively downregulated in L2 hypoxia compared with L4 hypoxia (B1/R1) ($p = 0.0127 \rightarrow 0.1908$).

To investigate the hypothesis that methylation sites are associated with stress-induced alternative transcriptional start sites within operons, MTase motif-free promoter regions sharing the same operon with MTase motif-containing promoter regions (op0 and op1 respectively) are distinctly labeled from MTase motif-free promoter regions in an operon with no or only motif of particular MTase (0 and 1 respectively). Chi-squared analysis revealed considerably decreased proportions of op0 for all 3 MTases in upregulated genes and mildly increased association with downregulated genes (Table S4).

The effects of methylation on gene expression were analyzed across each operon by using paired Student's *t*-test on the 3 following metrics. First, means of log2fc suggested that genes with MamB sites might have lower, albeit far from significant, expression during hypoxia (Table S5), Second, the proportion of the number of operons in which genes with MT sites have lower overall log2fc hypoxia vs control than those without MT sites is the highest for MamB (Table S6). Third, a CH clustering validation index showed that clustering genes within operon by the presence of MamB sites demonstrated a higher (non-significant in adjusted *p*-values) clustering performance than the median of all possible 2-group combinations (Table S7).

CDC1551 was chosen as a representative strain of L4 instead of H37Rv since all of their genes have been mapped into operons and that all 4 promoter-associated MamA sites described by Shell et al. (2013) can be located using the rule of at most 200 bp upstream sequence. However, in H37Rv, *whiB7* or *MT3290.1* is preceded by MT3291 since MT3290.2 is not annotated. This would imply the presence of a 5′UTR of 353 bp directly upstream of the first gene in the operon, specifically MT3290.1 in this case, which would be exceptionally long for a ribosome-binding site (reviewed in Sawyer et al. 2018) or particular secondary RNA structure including attenuators (Canestrari et al. 2020). One possible caveat is that CDC1551 is phylogenetically distant from H37Rv even though they belong to the same clade of L4 (Zhu et al. 2016). Nevertheless, Tizzano et al. (2021) have shown that Haarlem strain exhibited a growth/survival profile similar to that of H37Rv during hypoxia. The analysis of distribution of MT sites compared between H37Rv and CDC1441 will be included in the future.

Association analysis using chi-squared test revealed that MamB methylation has a slightly more significant effect on distribution of hypoxia differential expressed genes than the other 2 MTases in L2. In addition, the effect of MamB on distribution of differential expressed genes is also lower in L4. This is partially congruent with the lack of MamA function in L2 and lack of MamB function in several strains of L4. Nonetheless, the similar analysis results of one MTase between the functional strain and loss-of-function strain could be attributed to the remaining homologous motif sequences that have not been modified much throughout the evolution. To illustrate, L4 Mtb F1 and F28 still have 3,902 and 3,904 MamA sitee, almost as same as 3,902 sites in L2 Mtb 2279.

One hypothesis is that methylation site may be associated with alternative promoter within an operon such that the genes upstream and genes downstream to the promoter methylation site could be differentially regulated. Despite not statistically significant after *p*-value adjustment, methylation especially MamB shows potential effects on differential expression of genes in the same operon as indicated by paired *t*-test of mean log2fc and CH internal clustering validation index. The proportion of operons in which genes with methylation sites are downregulated compared with those without is also highest for MamB. However, similar results were also observed in the strain with loss-of-function in corresponding MTase. Furthermore, Shell et al. (2013) showed that *mamA* deletion decreases the expressions of the 4 genes with MamA-associated promoter site even in baseline normal condition. If the methylation is only involved in directing an alternative transcriptional start site, those genes downstream to the MamA site should still be transcribed along their upstream genes in the same operon. One possible scenario for this transcriptional uncoupling is that the methylation motif sequence may induce transcription termination or RNA polymerase dissociation unless it has been methylated.

# Conclusion

Although the direct data on the effects of DNA methylation on gene expression during hypoxia of MTB are still lacking, several pieces of evidence including the our current meta-analysis suggests a potential association between methylation sites and hypoxia-induced differential expression in both MTB lineages with completely different profiles of MTases. However, the results are still inconclusive. Whole genome methylation analysis in conjunction with RNA sequencing across different hypoxic conditions is necessary for a better understanding of gene expression during hypoxia.

# Appendix - Meta-analysis Results

**Table S1.** A total of 47 genes estimated to be differentially regulated in L2 compared with L4.

| | R1/R0 | B1/R1 | z.B0/R0 | B1/B0 | r.B/R | R1/R0.dosR |
|---|---|---|---|---|---|---|
| *Rv3862c* | 0.000 | 6.510 | 0.000 | 6.510 | 6.510 | 0.000 |
| *Rv0083* | -3.063 | 5.062 | 0.000 | 1.999 | 5.062 | 1.070 |
| *Rv3829c* | 0.000 | 3.823 | 0.000 | 3.823 | 3.823 | 0.000 |
| *Rv3877* | 0.000 | 3.513 | 0.000 | 3.513 | 3.513 | 0.000 |
| *Rv3861* | 0.000 | 3.213 | 0.000 | 3.213 | 3.213 | 0.000 |
| *Rv2660c* | -3.753 | 2.859 | 0.000 | -0.894 | 2.859 | 0.000 |
| *Rv0782* | 0.000 | 2.820 | 0.000 | 2.820 | 2.820 | 0.000 |
| *Rv2428* | 1.818 | 2.251 | 0.000 | 4.069 | 2.251 | 0.000 |
| *Rv1371* | -1.297 | 2.244 | 0.000 | 0.947 | 2.244 | 0.000 |
| *Rv0826* | 1.878 | 3.590 | 1.438 | 4.030 | 2.152 | 0.000 |
| *Rv0186A* | -2.132 | 2.130 | 0.000 | -0.002 | 2.130 | 0.000 |
| *Rv3182* | -1.781 | 2.074 | 0.000 | 0.293 | 2.074 | 0.000 |
| *Rv0847* | -1.575 | 1.866 | 0.000 | 0.291 | 1.866 | 0.000 |
| *Rv2429* | 2.427 | 1.857 | 0.000 | 4.284 | 1.857 | 0.000 |
| *Rv0169* | -2.993 | 1.710 | 0.000 | -1.283 | 1.710 | 0.000 |
| *Rv0972c* | 1.220 | 1.676 | 0.000 | 2.896 | 1.676 | 0.000 |
| *Rv0969* | 1.003 | 1.614 | 0.000 | 2.616 | 1.614 | 0.000 |
| *Rv3127* | 4.145 | 0.000 | 1.590 | 2.555 | -1.590 | 4.121 |
| *Rv3049c* | 1.291 | -1.620 | 0.000 | -0.329 | -1.620 | 0.000 |
| *Rv3130c* | 3.604 | 0.000 | 1.642 | 1.963 | -1.642 | 4.672 |
| *Rv2031c* | 3.901 | 0.000 | 1.705 | 2.197 | -1.705 | 4.802 |
| *Rv0886* | 1.860 | -1.752 | 0.000 | 0.108 | -1.752 | 0.000 |
| *Rv3581c* | -1.110 | -1.866 | 0.000 | -2.976 | -1.866 | 0.000 |
| *Rv2107* | 1.547 | 0.000 | 1.903 | -0.356 | -1.903 | 0.000 |
| *Rv2626c* | 4.080 | 0.000 | 1.911 | 2.170 | -1.911 | 4.615 |
| *Rv2623* | 3.379 | 0.000 | 2.029 | 1.350 | -2.029 | 4.233 |
| *Rv3131* | 3.417 | 0.000 | 2.195 | 1.222 | -2.195 | 5.092 |
| *Rv2160A* | -1.490 | -2.349 | 0.000 | -3.840 | -2.349 | 0.000 |
| *Rv2628* | 2.351 | 0.000 | 2.782 | -0.431 | -2.782 | 3.766 |
| *Rv2627c* | 3.734 | 0.000 | 2.903 | 0.831 | -2.903 | 3.632 |
| *Rv0071* | 0.000 | -3.223 | 0.000 | -3.223 | -3.223 | 0.000 |
| *Rv1581c* | 0.000 | -3.464 | 0.000 | -3.464 | -3.464 | 0.000 |
| *Rv2160c* | -2.044 | -3.517 | 0.000 | -5.561 | -3.517 | 0.000 |
| *Rv1578c* | 0.000 | -3.697 | 0.000 | -3.697 | -3.697 | 0.000 |
| *Rv1585c* | 0.000 | -3.796 | 0.000 | -3.796 | -3.796 | 0.000 |
| *Rv1587c* | 0.000 | -3.872 | 0.000 | -3.872 | -3.872 | 0.000 |
| *Rv1762c* | 0.000 | -4.374 | 0.000 | -4.374 | -4.374 | 0.000 |
| *Rv1579c* | 0.000 | -4.533 | 0.000 | -4.533 | -4.533 | 0.000 |
| *Rv1584c* | -1.252 | -4.830 | 0.000 | -6.082 | -4.830 | 0.000 |
| *Rv1761c* | 0.000 | -4.891 | 0.000 | -4.891 | -4.891 | 0.000 |
| *Rv0072* | 0.000 | -4.943 | 0.000 | -4.943 | -4.943 | 0.000 |
| *Rv1580c* | 0.000 | -5.103 | 0.000 | -5.103 | -5.103 | 0.000 |
| *Rv2817c* | 0.000 | -5.702 | 0.000 | -5.702 | -5.702 | 0.000 |
| *Rv2819c* | 0.000 | -5.751 | 0.000 | -5.751 | -5.751 | 0.000 |
| *Rv1576c* | 0.000 | -5.755 | 0.000 | -5.755 | -5.755 | 0.000 |
| *Rv2816c* | -1.041 | -6.123 | 0.000 | -7.164 | -6.123 | 0.000 |
| *Rv0073* | 0.000 | -7.954 | 0.000 | -7.954 | -7.954 | 0.000 |

R0 – L4 in normal condition; R1 – L4 in hypoxia; B0 – L2 in normal condition; B1 – L2 in hypoxia; r.B/R – ratio of B1/B0 to R1/R0.

**Table S2.** Names and descriptions of the genes provided above.

| | gene | description |
|---|---|---|
| *Rv3862c* | *whiB6* | Possible transcriptional regulatory protein WhiB-like WhiB6 NP_218379.1 [4338170:4338521] -1 |
| *Rv0083* | | Probable oxidoreductase NP_214597.1 [90399:92322] 1 |
| *Rv3829c* | | Probable dehydrogenase NP_218346.1 [4303397:4305008] -1 |
| *Rv3877* | *eccD1* | ESX conserved component EccD1 ESX-1 type VII secretion system protein Probable transmembrane protein NP_218394.1 [4355006:4356542] 1 |
| *Rv3861* | | Hypothetical protein NP_218378.1 [4337945:4338272] 1 |
| *Rv2660c* | | Hypothetical protein NP_217176.1 [2980962:2981190] -1 |
| *Rv0782* | *ptrBb* | Probable protease II PtrBb [second part] (oligopeptidase B) NP_215295.2 [874731:876390] 1 |
| *Rv2428* | *ahpC* | Alkyl hydroperoxide reductase C protein AhpC (alkyl hydroperoxidase C) NP_216944.1 [2726192:2726780] 1 |
| *Rv1371* | | Probable conserved membrane protein NP_215887.1 [1543358:1544828] 1 |
| *Rv0826* | | hypothetical protein NP_215341.1 [919633:920689] 1 |
| *Rv0186A* | *mymT* | Metallothionein, MymT YP_004837046.2 [218389:218551] -1 |
| *Rv3182* | | hypothetical protein NP_217698.1 [3550373:3550718] 1 |
| *Rv0847* | *lpqS* | Probable lipoprotein LpqS NP_215362.1 [944342:944735] 1 |
| *Rv2429* | *ahpD* | Alkyl hydroperoxide reductase D protein AhpD (alkyl hydroperoxidase D) NP_216945.1 [2726805:2727339] 1 |
| *Rv0169* | *mce1A* | Mce-family protein Mce1A YP_177701.1 [198533:199898] 1 |
| *Rv0972c* | *fadE12* | Acyl-CoA dehydrogenase FadE12 NP_215487.1 [1082583:1083750] -1 |
| *Rv0969* | *ctpV* | Probable metal cation transporter P-type ATPase CtpV NP_215484.3 [1078742:1081055] 1 |
| *Rv3127* | | hypothetical protein NP_217643.1 [3492146:3493181] 1 |
| *Rv3049c* | | Probable monooxygenase NP_217565.1 [3409508:3411083] -1 |
| *Rv3130c* | *tgs1* | Triacylglycerol synthase (diacylglycerol acyltransferase) Tgs1 NP_217646.1 [3494974:3496366] -1 |
| *Rv2031c* | *hspX* | Heat shock protein HspX (alpha-crystallin homolog) (14 kDa antigen) (HSP16.3) NP_216547.1 [2278497:2278932] -1 |
| *Rv0886* | *fprB* | Probable NADPH:adrenodoxin oxidoreductase FprB (adrenodoxin reductase) (AR) (ferredoxin-NADP(+) reductase) NP_215401.1 [983802:985530] 1 |
| *Rv3581c* | *ispF* | Probable 2C-methyl-D-erythritol 2,4-cyclodiphosphate synthase IspF (MECPS) NP_218098.1 [4023867:4024347] -1 |
| *Rv2107* | *PE22* | PE family protein PE22 YP_177858.1 [2367358:2367655] 1 |
| *Rv2626c* | *hrp1* | Hypoxic response protein 1 Hrp1 NP_217142.1 [2952561:2952993] -1 |
| *Rv2623* | *TB31.7* | Universal stress protein family protein TB31.7 NP_217139.1 [2949592:2950486] 1 |
| *Rv3131* | | hypothetical protein NP_217647.3 [3496550:3497549] 1 |
| *Rv2160A* | | hypothetical protein YP_177660.1 [2421642:2422278] -1 |
| *Rv2628* | | Hypothetical protein NP_217144.1 [2955057:2955420] 1 |
| *Rv2627c* | | hypothetical protein NP_217143.1 [2953506:2954748] -1 |
| *Rv0071* | | Possible maturase NP_214585.1 [79485:80193] 1 |
| *Rv1581c* | | Probable PhiRv1 phage protein NP_216097.1 [1783905:1784301] -1 |
| *Rv2160c* | | hypothetical protein NP_216676.1 [2421661:2422003] -1 |
| *Rv1578c* | | Probable PhiRv1 phage protein NP_216094.1 [1782757:1783228] -1 |
| *Rv1585c* | | Possible phage PhiRv1 protein NP_216101.1 [1786583:1787099] -1 |
| *Rv1587c* | | Partial REP13E12 repeat protein NP_216103.2 [1788161:1789163] -1 |
| *Rv1762c* | | hypothetical protein NP_216278.1 [1995053:1995842] -1 |
| *Rv1579c* | | Probable PhiRv1 phage protein NP_216095.1 [1783308:1783623] -1 |
| *Rv1584c* | | Possible PhiRv1 phage protein NP_216100.1 [1786306:1786528] -1 |
| *Rv1761c* | | Possible exported protein NP_216277.1 [1994670:1995054] -1 |
| *Rv0072* | | Probable glutamine-transport transmembrane protein ABC transporter NP_214586.1 [80623:81673] 1 |
| *Rv1580c* | | Probable PhiRv1 phage protein NP_216096.1 [1783619:1783892] -1 |
| *Rv2817c* | | hypothetical protein NP_217333.1 [3123966:3124983] -1 |
| *Rv2819c* | | Hypothetical protein NP_217335.1 [3126239:3127367] -1 |
| *Rv1576c* | | Probable PhiRv1 phage protein NP_216092.1 [1780642:1782064] -1 |
| *Rv2816c* | | hypothetical protein NP_217332.1 [3123624:3123966] -1 |
| *Rv0073* | | Probable glutamine-transport ATP-binding protein ABC transporter NP_214587.1 [81675:82668] 1 |

1 main strand; -1 complementary strand

**Table S3.** Chi-squared analysis between genes with (1) or without (0) promoter-associated MT-sites and differentially/regulated genes in given condition

| Genome | MTase | Condition | $\chi^2$-test $p$-value | Important residuals |
|---|---|---|---|---|
| L2 | MamA | B1/B0 | 0.0067 | 1-down (-2.2549) |
| L2 | MamA | B1/R1 | 0.0806 | 1-up (2.2398) |
| L2 | MamB | B1/R1 | 0.0805 | 1-up (-1.6189) |
| L4 | MamA | R1/R0 | 0.0050 | 1-down (-2.2206) |
| L4 | MamB | B1/R1 | 0.0127 | 1-down (1.9882); 1-up (-1.8180) |
| L4 | MamC | R1/R0 | 0.0900 | 1-up (-1.9832) |

**Table S4.** Chi-squared analysis between genes with (op1) or without (op0) promoter MT-sites sharing the same operon or genes independently with (1) or without (0) MT sites and differentially/regulated genes.

| Genome | MTase | Condition | $\chi^2$-test $p$-value | Important residuals |
|---|---|---|---|---|
| L2 | MamA | B1/B0 | $0.0292^+$ | op0-up (-2.8088); 0-up (1.5050) |
| L2 | MamB | B1/B0 | 0.0022* | op0-up (-3.5661); 0-up (1.7089) |
| L2 | MamC | B1/B0 | $0.0263^+$ | op0-up (-2.7409); 0-up (1.4979) |
| L2 | MamA | r.B/R | 0.0148* | op0-up (2.3681); 0-up (-1.8095) |
| L2 | MamB | r.B/R | 0.0617 | op0-up (2.4895); 0-up (-1.3915) |
| L2 | MamC | r.B/R | 0.0074* | op0-up (3.5507); 0-up (-1.6101) |
| L4 | MamA | R1/R0 | < 0.0001*** | op0-up (-3.8063); 0-up (2.3365) |
| L4 | MamB | R1/R0 | 0.0001*** | op0-up (-3.8497); 1-down (2.2925) |
| L4 | MamC | R1/R0 | < 0.0001*** | op0-up (-4.4560); 0-up (2.4704) |
| L4 | MamA | r.B/R | 0.0092* | op0-up (2.4665); op1-up (2.2730) |
| L4 | MamB | r.B/R | 0.0729 | op0-up (2.7041); 0-up (-1.3840) |
| L4 | MamC | r.B/R | 0.0043* | op0-up (3.5592); 1-down (1.6180) |

$^+p < 0.10$; $*p < 0.05$; $***p < 0.001$ when adjusted by FDR.

**Table S5.** Paired $t$-test between means of genes with or without MT sites across the same operons

| Genome | MTase | Condition | N | Estimates | $p$-value |
|---|---|---|---|---|---|
| L2 | HsdM | B1/R1, r.B/R | 5 | 0.1110 | 0.1901 |
| L4 | HsdM | B1/R1, r.B/R | 7 | 0.1567 | 0.0806 |
| L2 | MamB | B1/B0 | 10 | -0.3115 | 0.5043 |
| L4 | MamB | R1/R0 | 9 | -0.4016 | 0.4957 |

**Table S6.** The numbers of operon with overall log2fc negative (Down), close to zeros (Neutral), or positive (Up) in given conditions

| Genome | MTase | Condition | Down | Neutral | Up |
|---|---|---|---|---|---|
| L2 | MamA | R1/R0 | 5 | 11 | 5 |
| L2 | MamB | R1/R0 | 4 | 5 | 1 |
| L2 | HsdM | R1/R0 | 2 | 1 | 2 |
| L2 | MamA | r.B/R | 1 | 19 | 1 |
| L2 | MamB | r.B/R | 0 | 10 | 0 |
| L2 | HsdM | r.B/R | 0 | 3 | 2 |
| L4 | MamA | R1/R0 | 6 | 6 | 5 |
| L4 | MamB | R1/R0 | 4 | 3 | 2 |
| L4 | HsdM | R1/R0 | 3 | 2 | 2 |
| L4 | MamA | r.B/R | 1 | 16 | 0 |
| L4 | MamB | r.B/R | 0 | 9 | 0 |
| L4 | HsdM | r.B/R | 0 | 4 | 3 |

**Table S7.** Paired *t*-test between means of CH indices of gene clustering by presence or absence of promoter-associated MT sites across the same operons

| Genome | MTase | Condition | N | CH Estimates | *p*-value |
|---|---|---|---|---|---|
| L4 | MamB | R1/R0 | 4 | 10.9248 | 0.0410 |
| L2 | MamB | B1/B0 | 6 | 10.5164 | 0.0622 |
| L2 | MamA | B1/B0 | 8 | 1.2922 | 0.1050 |
| L4 | MamA | R1/R0 | 9 | 1.0820 | 0.4214 |